\documentclass[usenatbib,useAMS]{mn2e}
\usepackage{natbib}
\usepackage{amsmath}
\usepackage{graphicx}
\usepackage{color}
\usepackage{epsfig}
\usepackage{comment}
\allowdisplaybreaks[1]
\usepackage{here}

\newcommand{\msun}{{\rm M}_\odot}

\newcommand{\cc}{{\rm cm}^{-3}}

\newcommand{\K}{{\rm K}}
\newcommand{\beq}{\begin{equation}}
\newcommand{\eeq}{\end{equation}}

\defcitealias{inayoshi+16}{IHO16}

\title[2D Hyper-Eddington accretion]
{
Rapid growth of black holes accompanied with hot or warm \\
outflows exposed to anisotropic super-Eddington radiation
}
\author[]{Eishun Takeo$^{1}$\thanks{takeo@kusastro.kyoto-u.ac.jp},
Kohei Inayoshi$^{2}$\thanks{inayoshi@astro.columbia.edu; Simons Society of Fellows (KI).},
Ken Ohsuga$^{3}$,
Hiroyuki R. Takahashi$^{3}$ and
\newauthor
Shin Mineshige$^{1}$
\\
$^{1}$Department of Astronomy, Graduate School of Science, Kyoto University, 
Kitashirakawa, Oiwakecho, Sakyo-ku, Kyoto 606-8502\\
$^{2}$Department of Astronomy, Columbia University, 550 W. 120th Street, New York, NY 10027, USA\\
$^{3}$National Astronomical Observatory of Japan, Osawa, Mitaka, Tokyo 181-8588, Japan
}

\begin{document}
\maketitle
\label{firstpage}

\begin{abstract}
We perform two-dimensional radiation hydrodynamical simulations of accretion flows
onto a black hole (BH) with a mass of $10^3\leq M_{\rm BH}/\msun \la 10^6$ 
in order to study rapid growth of BHs in the early Universe.
For spherically symmetric flows, hyper-Eddington accretion onto the BH from outside the Bondi radius
can occur unimpeded by radiation feedback only when the BH mass is higher than 
$\simeq 10^4~\msun(n_\infty/10^5~\cc)^{-1}(T_\infty/10^4~\K)^{3/2}$,
where $n_\infty$ and $T_\infty$ are the density and temperature of ambient gas.
Here, we study the properties of accretion flows exposed to anisotropic radiation
from a nuclear accretion disk with a luminosity higher than the Eddington value ($L_{\rm Edd}$)
due to collimation toward the bipolar directions.
We find that,
unlike the spherically symmetric case, even less massive 
BHs with $M_{\rm BH} < 10^4~\msun$ can be fed by surrounding gas 
at high accretion rates of $\ga L_{\rm Edd}/c^2$ through the equatorial plane,
while ionized regions expand to the polar directions producing hot outflows with $T\sim 10^5~\K$.
For more massive BHs with $M_{\rm BH}\ga 5\times 10^5~\msun$, 
neutral gas through the equatorial plane totally covers the central radiating region 
due to the non-radial gas motions, and thus the emergent radiation in all directions is blocked.
Because of efficient recombination by hydrogen, the entire flow results in neutral and warm gas with
$T \simeq 8000$ K.
The central BH is fed through the equator at the averaged rate of $\sim 5\times 10^4~L_{\rm Edd}/c^2$,
which corresponds to $\sim 50~\%$ of the inflow rate from the Bondi radius.
Moreover, radiation momentum absorbed by neutral hydrogen produces warm outflows toward the bipolar directions
at $\sim 30~\%$ of the BH feeding rate and with a typical velocity of $\simeq 50~{\rm km~s}^{-1}$.
\end{abstract}

\begin{keywords}
black hole physics, cosmology: theory, quasars: supermassive black holes
\end{keywords}


\section{Introduction}

Observations of bright quasars at high redshifts $(z\ga 6)$ have revealed that 
supermassive black holes (SMBHs) with $M_{\rm BH}\ga 10^9~\msun$ are formed within $\la 1$ Gyr 
after the beginning of the Universe \citep[e.g.][]{Fan_2004,Mortlock_2011,Wu_2015}.
Although SMBHs play crucial roles in the co-evolution with their host galaxies 
through feedback processes
\citep[e.g.][]{Silk_Rees_1998,King_2003,Murray_2005, Kormendy_Ho_2013}, 
their formation and growth processes are still open questions.

One possible formation scenario of high-$z$ SMBHs is that stellar mass black holes (BHs) 
formed by collapse of first generation stars (Population III, hereafter PopIII)
with $\sim 100~\msun$ grow via gas accretion \citep[e.g.,][]{MadauRees01,HaimanLoeb01,VHM03}.
In the standard accretion physics, BH growth would be suppressed by radiation emitted 
from a nuclear accretion disk 
because the radiation luminosity increases with the accretion rate onto the BH.
If the radiative efficiency of the disk is $\eta \simeq 0.1$ \citep{SS_1973,Soltan_1982,Yu_Tremaine_2002},
the radiation luminosity exceeds the Eddington luminosity ($L_{\rm Edd}$) 
at the accretion rate of $\ga L_{\rm Edd}/(\eta c^2)$, above which the radiation force dominates 
the gravity of the central BH.
Assuming this Eddington-limited accretion rate, it takes $\ga 1$ Gyr for stellar-mass
seed BHs to grow up to $\sim 10^9~\msun$. 
This fact requires more rapid accretion mechanisms to form SMBHs by $z\sim 6$, when the corresponding 
cosmic age is $\sim 0.8$ Gyr.

As alternative possibilities, formation of massive seed BHs with $\ga 10^3-10^5~\msun$ has been proposed:
direct-collapse of supermassive stars in protogalaxies 
\citep[e.g.,][]{2014MNRAS.439.1160R,IOT14,2015MNRAS.446.2380B,IT_2015,
2014MNRAS.445.1056V,Latif_2016,Chon_2016}
or runaway collisions of stars in a dense cluster
\citep[e.g.,][]{2009ApJ...694..302D,2015MNRAS.451.2352K,2016MNRAS.457.2423Y,Stone_2017,Sakurai_2017}.
However, since the BH growth time even from such massive seeds would be shorten only by a factor of two
in the case of the Eddington limited accretion, a high duty cycle ($\sim 100\%$) is still required to explain 
the existence of SMBHs with $\sim 10^9~\msun$ at high redshifts.

The possibility of rapid growth of seed BHs has been discussed
\citep[e.g.][]{VR_2005,TH09,Alexander_2014,Madau_2014,2015MNRAS.448..104P,Valiante_2016,Pezzulli_2016}.
By means of radiation hydrodynamical simulations of accretion flows onto a BH,
\cite{ohsuga+05} concluded that 
super-Eddington accretion can be realized as long as sufficient gas is 
supplied at the vicinity of the central BH 
\citep[see also][]{ohsuga+2009,ohsuga_mineshige2011,Jiang_2014,Sadowski_2015,takahashi2016}.
This is because diffusive photons are efficiently trapped within optically-thick accretion flows
and the radiation force onto the gas is alleviated \citep[e.g.][]{Begelman_1979}.
However, it is unclear how sufficient amount of gas is supplied into the nuclear region.
In fact, radiation heating and ionization effectively suppress gas accretion from larger scales
\citep{Ciotti_Ostriker_2001,Milosavljevic_2009,Alvarez_2009,2009ApJ...699...89C,Park_Ricotti_2011,Park_Ricotti_2012}.

\cite{inayoshi+16} (hereafter \citetalias{inayoshi+16}) 
have found a self-consistent solution of steady hyper-Eddington accretion 
in spherically symmetric systems with radiation hydrodynamical simulations.
This accretion flow has a high accretion rate of $\ga 5000~L_{\rm Edd}/c^2$, 
unimpeded by negative feedback due to radiation force and heating simultaneously.
The solution is composed of a radiation-dominated core, where photon trapping due to electron 
scattering works efficiently, and an accreting envelope which follows a Bondi profile with $8000~\K$. 
When the emergent radiation luminosity is limited to $\la L_{\rm Edd}$ because of photon trapping, 
radiation from the central region does not affect the gas dynamics at larger scales. 
The hyper-Eddington accretion is then realized when a BH is embedded in a dense gas cloud,
for which the following condition is satisfied,
\begin{equation}
M_{\rm BH} \ga 10^4~\msun 
\left(\frac{n_\infty}{10^5~\cc}\right)^{-1}\left(\frac{T_\infty}{10^4~\K}\right)^{3/2},
\label{eq:Hyper_Edd}
\end{equation}
where $n_\infty$ and $T_\infty$ are density and temperature of the ambient gas, respectively.
Hereafter, we define $M_{{\rm BH},x}=(M_{\rm BH}/10^x~\msun)$, 
$n_{\infty,5}=(n_\infty/10^5~\cc)$, and
$T_{\infty,4}=(T_\infty/10^4~\K)$.
Note that Eq.(\ref{eq:Hyper_Edd}) is identical to the condition where the Bondi radius is larger than
the size of the ionized region.
Even if a high luminosity with $>L_{\rm Edd}$ emerges from the center, the above conditions are not affected
as long as the luminosity is as low as $(10-30)\times L_{\rm Edd}$ \citep{Sakurai_2016}.
To satisfy the condition of Eq. (\ref{eq:Hyper_Edd}), gas-rich regions with $n\ga 10^7~\cc$ are required
for stellar-mass PopIII BHs with masses of $\la 100~\msun$.
During the assembly of a protogalaxy, some PopIII BHs would fall into a gaseous circumnuclear disk,
where the gas accretion rate is likely to be $\ga 5000~L_{\rm Edd}/c^2$ (\citealt{Ryu_2016}, see also 
\citealt{Lupi_2016}).

Radiation flux from an accretion disk around the nuclear BH is more likely to be anisotropic,
i.e., preferentially collimated around the rotation axis of the disk
\citep{ohsuga+2009,ohsuga_mineshige2011,Jiang_2014,Sadowski_2015,takahashi2016}.
On the other hand, the disk-like accretion could reduce the feedback effect
because intense radiation is significantly attenuated by the dense accreting material in the disk
\citep{ohsuga_mineshige_2007,sugimura+16}.
In this paper, we investigate how intermediate massive BHs with $10^3\leq M_{\rm BH}\leq 5 \times 10^5 ~\msun$, 
embedded in an initially uniform gas with $n_\infty=10^5~\cc$ and $T_\infty =10^4~\K$,
can grow via accretion when the inflow gas is exposed to the super-Eddington, anisotropic radiation.
To address this issue, we perform two-dimensional radiation hydrodynamical simulations,
including multifrequency radiation transfer and non-equilibrium primordial chemistry.
We confirm that hyper-Eddington accretion from the Bondi radius is likely to occur 
when radiation is more anisotropic and the BH mass is higher.
Furthermore, we demonstrate that for $M_{\rm BH} \ga 5\times 10^5~\msun$,
neutral gas inflows through the equator totally cover the central radiating region
due to the non-radial gas motions, and thus the emergent radiation in all directions is blocked. 
Because of efficient recombination by hydrogen, the entire flow results in neutral and warm gas.
In this case, the BH feeding rate is as high as the Bondi accretion rate, and also warm outflows are produced 
by absorption of radiation momentum.

The rest of this paper is organized as follows. 
In Section 2, we describe the methodology for radiation hydrodynamical simulations. 
In Section 3, we show our simulation results and discuss the effects of anisotropic radiation.
Finally, we summarize the main conclusions of this paper and 
discuss caveats of our simulations in Section 4.


\section{Simulation method}

We perform two-dimensional hydrodynamical simulations including one-dimensional 
radiation transfer and chemical reaction networks.
Our purpose is to study necessary conditions
for hyper-Eddington accretion led by sufficient supply of gas from large scales.
Gas accretion begins from the Bondi radius, within which the gravity of the central BH
dominates the gas-pressure gradient force, given by \cite{Bondi_1952}
\begin{equation}
R_{\rm B}\equiv \frac{GM_{\rm BH}}{c_{\infty}^2} \simeq 1.97\times10^{18}
~M_{\rm BH,4}
~T_{\infty,4}^{-1} \ {\rm cm},
\end{equation}
where $c_{\infty}=\sqrt{\gamma \mathcal{R} T_{\infty} /\bar{\mu}}$ is the sound speed,
$\gamma$ is the specific heat ratio, $\mathcal{R}$ is the gas constant, 
and $\bar{\mu}$ is the mean molecular weight. 
As a reference of the accretion rate, we define the Bondi accretion rate for $\gamma=1$ as
\begin{equation}
  \dot{M}_{\rm B} \equiv \pi e^{3/2} \rho_\infty \frac{G^2 M_{\rm BH}^2}{c_{\infty}^3},
\end{equation}
and the Eddington accretion rate as $\dot{M}_{\rm Edd} \equiv L_{\rm Edd}/c^2$,
where $L_{\rm Edd} = 4\pi c GM_{\rm BH}/\kappa_{\rm es}$ and $\kappa_{\rm es}$ is the electron scattering opacity.
Note that the Bondi radius and rate are calculated for $\gamma=1$, $\bar{\mu}=1.23$ and 
$T_{\infty}=10^4 {\rm K}$ as a reference value through this paper, and that these are self-consistently calculated in our simulations.

\subsection{Basic equations}
\label{sec:BE}
We consider a gas sphere exposed to intense radiation from the accretion flow 
onto the central BH. 
We adopt the spherical coordinates of $(r, \theta, \phi)$,
defining the polar axis ($\theta = 0$ and $\pi$) as directions 
perpendicular to the disk plane.

We solve two-dimensional hydrodynamical equations assuming axisymmetric flows;
the equation of continuity
\begin{equation}
  \frac{\partial \rho}{\partial t} + \nabla\cdot(\rho {\boldsymbol v}) = 0,
  \label{renzoku}
\end{equation}
and the equations of motion,
\begin{equation}
  \frac{\partial \left(\rho v_r\right)}{\partial t} + \nabla\cdot(\rho v_r {\boldsymbol v}) 
  = -\frac{\partial p}{\partial r} + \rho\left( \frac{v_{\theta}^2}{r} +\frac{v_{\phi}^2}{r} \right) - \rho \frac{\partial \psi}{\partial r}+ f_{\rm rad},
  \label{eomr}
\end{equation}
\begin{equation}
  \frac{\partial \left(\rho rv_{\theta}\right)}{\partial t} + \nabla\cdot(\rho r v_{\theta} {\boldsymbol v}) 
  = -\frac{\partial p}{\partial \theta} + \rho v_{\phi}^2\cot{\theta},
  \label{eomt}
\end{equation}
\begin{equation}
  \frac{\partial \left(\rho rv_{\phi}\sin{\theta}\right)}{\partial t} + \nabla\cdot\left(\rho r v_{\phi}\sin{\theta} {\boldsymbol v}\right) = 0,
  \label{eomp}
\end{equation}
where $\rho$ is the gas density, ${\boldsymbol v}=(v_r,v_{\theta}, v_{\phi})$ is the velocity, 
$p$ is the gas pressure, and $f_{\rm rad}$ is the radiation force.
We consider the gravity of the central BH ($r=0$) and neglect the gas self-gravity. 
Since the general relativistic effect is negligible, the gravitational potential is given by 
$\psi=-GM_{\rm BH}/r$.

We also solve the energy equation 
\begin{equation}
  \frac{\partial e}{\partial t} + \nabla\cdot[(e+p){\boldsymbol v}] = -\frac{GM_{\rm BH}\rho}{r^2}v_r -\Lambda + \Gamma,
  \label{eneeq}
\end{equation}
where $e=e_{\rm int}+\rho |{\boldsymbol v}|^2/2$ is the gas total energy density, $e_{\rm int}$ is the gas internal energy density, 
$\Lambda$ is the net cooling rate per volume and $\Gamma$ is the heating rate due to radiation from the central region.
We assume the equation of state of ideal gas as $p=(\gamma-1)e_{\rm int}$ for $\gamma=5/3$.

We consider radiative cooling by bound-bound transitions of ${\rm H, He, He^+}$ 
atoms and free-free emission \citep{glover+07}. 
To estimate their rates, we solve chemical reaction networks including six species of 
H, H$^+$, He, He$^+$, He$^{++}$, and ${\rm e^-}$. 
The abundance of He nuclei relative to H nuclei is set to $8.33\times10^{-2}$.
Here, photoionization, collisional ionization and radiative recombination are considered \citep{abel+97,glover+07}. 
Instead of treating photoionization by diffusive recombination photons, we adopt the on-the-spot approximation, 
where the case A recombination rate is replaced by the case B rate.
To update the chemical abundances stably, we adopt a semi-implicit method \citep{anninos+97},
setting time steps shorter than chemical timescales defined by
$t_{\rm chem} \equiv (x_{\rm e}+0.001x_{\rm H})/\dot{x}_{\rm e}$,
where $x_{\rm e}$ and $x_{\rm H}$ are the abundance of electrons and neutral hydrogens, respectively
\citep{Whalen_2006,Whalen_2008}.

We solve the multi-frequency, steady radiative transfer equation
because the light crossing time is much shorter than the hydrodynamical timescale,
\begin{equation}
\frac{1}{r^2}\frac{{\rm d}}{{\rm d}r}(r^2F_{\nu}) = -\rho\kappa_{\nu}cE_{\nu},
\label{rteq}
\end{equation}
where $F_{\nu}$ is the radiation flux, $E_{\nu}$ is the radiation energy density, and 
$\kappa_{\nu}$ is the absorption opacity.
Note that the radial component of the radiation flux is calculated 
(see \S\ref{sec:discuss} for more details).
Since the ionized gas is optically thin to electron scattering and bound-free transitions,
we assume $F_{\nu} \approx cE_{\nu}$ on the right-hand-side of Eq. (\ref{rteq}).
The frequency range is set to $h\nu_{\rm min}(=13.6~{\rm eV}) \leq h\nu \leq h\nu_{\rm max}(=10~{\rm keV})$,
where $h$ is the Planck constant.
The ionization rate coefficients $k_{\rm ph}$ of ${\rm H, He, He^+}$ and photoionization heating rate 
$\Gamma(=\Gamma_{\rm H})$
are calculated 
following the photon-conserving manner \citep{Whalen_2006}.
Here $\Gamma_{\rm H}$ is the heating rate due to H ionization.
We consider radiation force due to electron scattering and bound-free absorption of H atoms as
\begin{equation}
  f_{\rm rad} = \frac{nx_{\rm e}}{c}\int_{\nu_{\rm min}}^{\nu_{\rm max}}\sigma_{\rm es}F_{\nu}{\rm d}\nu + \frac{\Gamma_{\rm H}}{c}.
  \label{f_rad}
\end{equation}
%


\begin{table}
\begin{center}
\begin{tabular}{ccccc}
Model & $M_{\rm BH}~(\msun)$ & $N$ & $\dot{M}/\dot{M}_{\rm Edd}$ & $L /L_{\rm Edd}$\\
\hline 
A & $10^3$ & 0& $\sim 0.9$ & $\sim 0.09$\\
B & $10^3$ & 2& $\approx 2.7$ & $\approx 0.27$\\
C & $10^3$ & 4 &$\approx 44$ & $\approx 3.5$\\
D & $5 \times 10^4$ & 4 &$\sim 1 \times 10^3 $ & $\sim 10 $\\
E & $5 \times 10^5$ & 4 & $\sim 5 \times 10^4  $ & $\sim 18 $\\
\hline
\end{tabular}
\end{center}
Column (1) model ID, (2) BH mass, (3) radiation anisotropy, namely $F_{\rm rad}(\theta)\propto \cos^N \theta$,
(4) accretion rate and (5) corresponding luminosity.
In column 4 and 5, we show the time-averaged values in Model A,
the values at the end of simulations in Models B-D, and
the time-averaged values over the last one cycle of episodic accretion in Model E.
\label{models}
\end{table}

In order to integrate Eq. (\ref{rteq}), we simply set a radiation source with a single power-law spectrum at the center,
$L_{\nu} = L_0  (\nu/ \nu_{\rm min})^{-\alpha}$ at $\nu_{\rm min} \leq \nu \leq \nu_{\rm max}$
and $L_{\nu}=0$ at $\nu \leq \nu_{\rm min}$ and $\nu_{\rm max} \leq \nu$, where $\alpha = 1.5$ is adopted.
The normalization factor $L_0$ is set by the total luminosity ($L=\int_{\nu_{\rm min}}^{\nu_{\rm max}} L_{\nu} {\rm d}\nu$).
We set a model for radiation luminosity emitted by an accretion disk, which is unresolved,
at the inner boundary,
\begin{equation}
  \frac{\displaystyle L}{\displaystyle L_{\rm Edd}}= \left\{ \begin{array}{ll}
    2\left[1+\ln{(\dot{m}/20)}\right] & {\rm for}\ \dot{m}\geq20, \\
    \dot{m}/10 & {\rm for}\ \dot{m}<20,
  \end{array}
  \right.
  \label{lumi}
\end{equation}
\citep{watarai+00}, where $\dot{m} \equiv \dot{M}/\dot{M}_{\rm Edd}$.
Furthermore, we assume anisotropic radiation fields as 
\begin{equation}
  F_{\nu}(r=r_{\rm min},\theta) = \frac{\displaystyle (N+1)L_{\nu}}
  {\displaystyle 4\pi r_{\rm min}^2}\cos^N{\theta},
  \label{frad_in}
\end{equation}
where the radiation flux $F_{\nu}(r_{\rm min})$ is normalized so that 
$L_{\nu}(r_{\rm min})=\int F_{\nu}(r_{\rm min}) r_{\rm min}^2 {\rm d}\Omega$
and $r_{\rm min}$ is the radius of the inner-most grid in the computation domain (see \S\ref{sec:IBC}).
Unlike \cite{sugimura+16}, shadow regions are not assumed in our simulations
(see \S\ref{sec:discuss} for more details).
Here, we explore the effects of radiation anisotropy;
the isotropic case $N = 0$ (Model A), the anisotropic cases $N=2$ (Model B), and $N=4$ (Model C) 
for $M_{\rm BH}=10^3~\msun$.
We also study cases for higher BH masses with 
$M_{\rm BH}=5 \times 10^4~\msun$ (Model D) 
and $5\times 10^5~\msun$ (Model E) for $N=4$.
Our models are summarized in Table 1.

\begin{figure}
\begin{center}
\includegraphics[width=8.4cm]{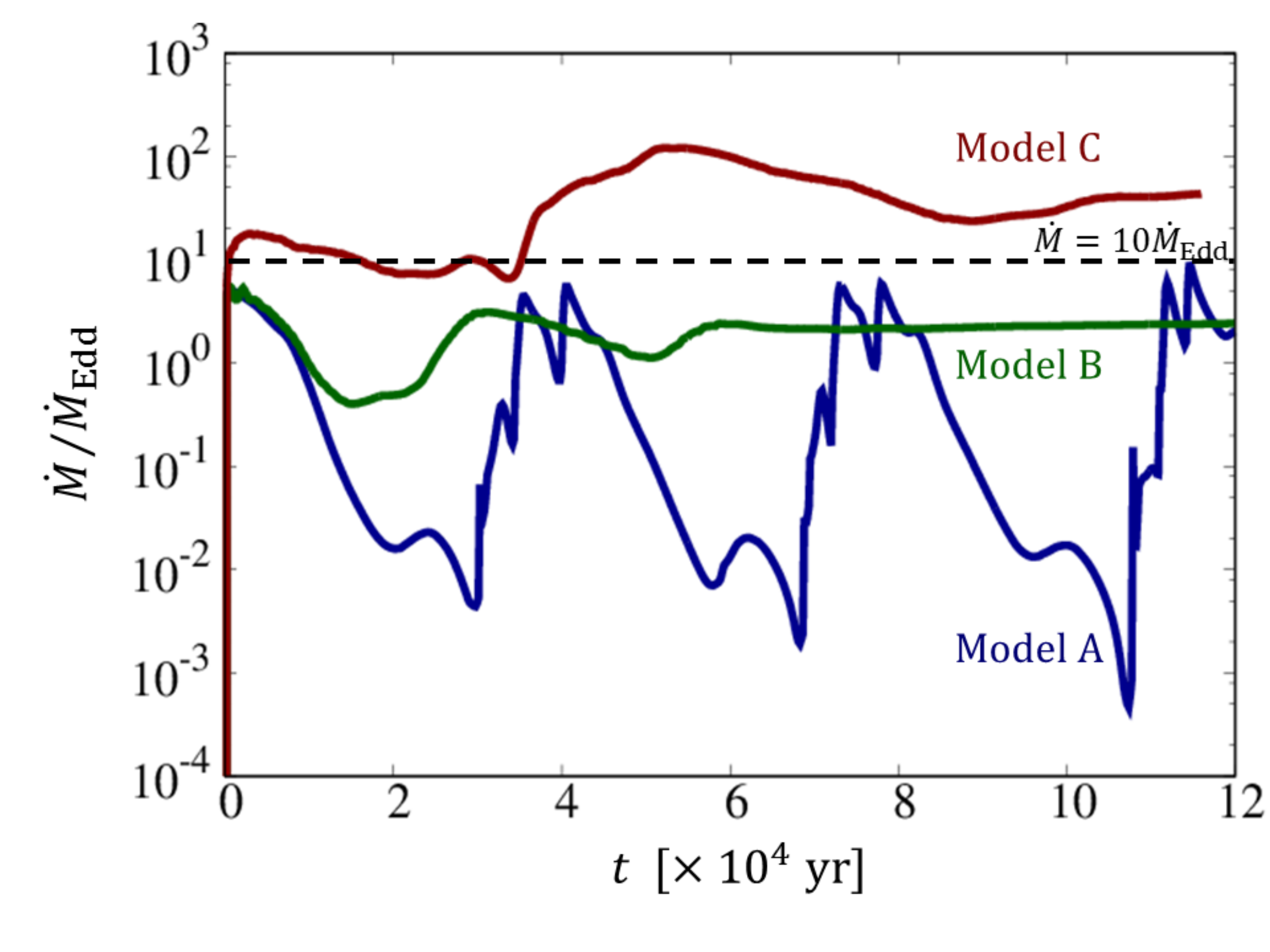}
\end{center}
\vspace{-4mm}
\caption{Time evolution of gas accretion rates onto a $10^3~\msun$ BH
in cases with isotropic radiation (Model A) and anisotropic radiation, $F_{\rm rad} \propto \cos^2{\theta}$ (Model B) and 
$F_{\rm rad} \propto \cos^4{\theta}$ (Model C). 
In Model A, the accretion occurs episodically and the time-averaged rate is below $\dot{M}_{\rm Edd}$.
The accretion rate in Model C exceeds the Eddington rate ($\dot{M}>10\dot{M}_{\rm Edd}$).
}
\label{time-lmd}
\end{figure}

\begin{figure*}
\begin{center}
\includegraphics[width=13.5cm]{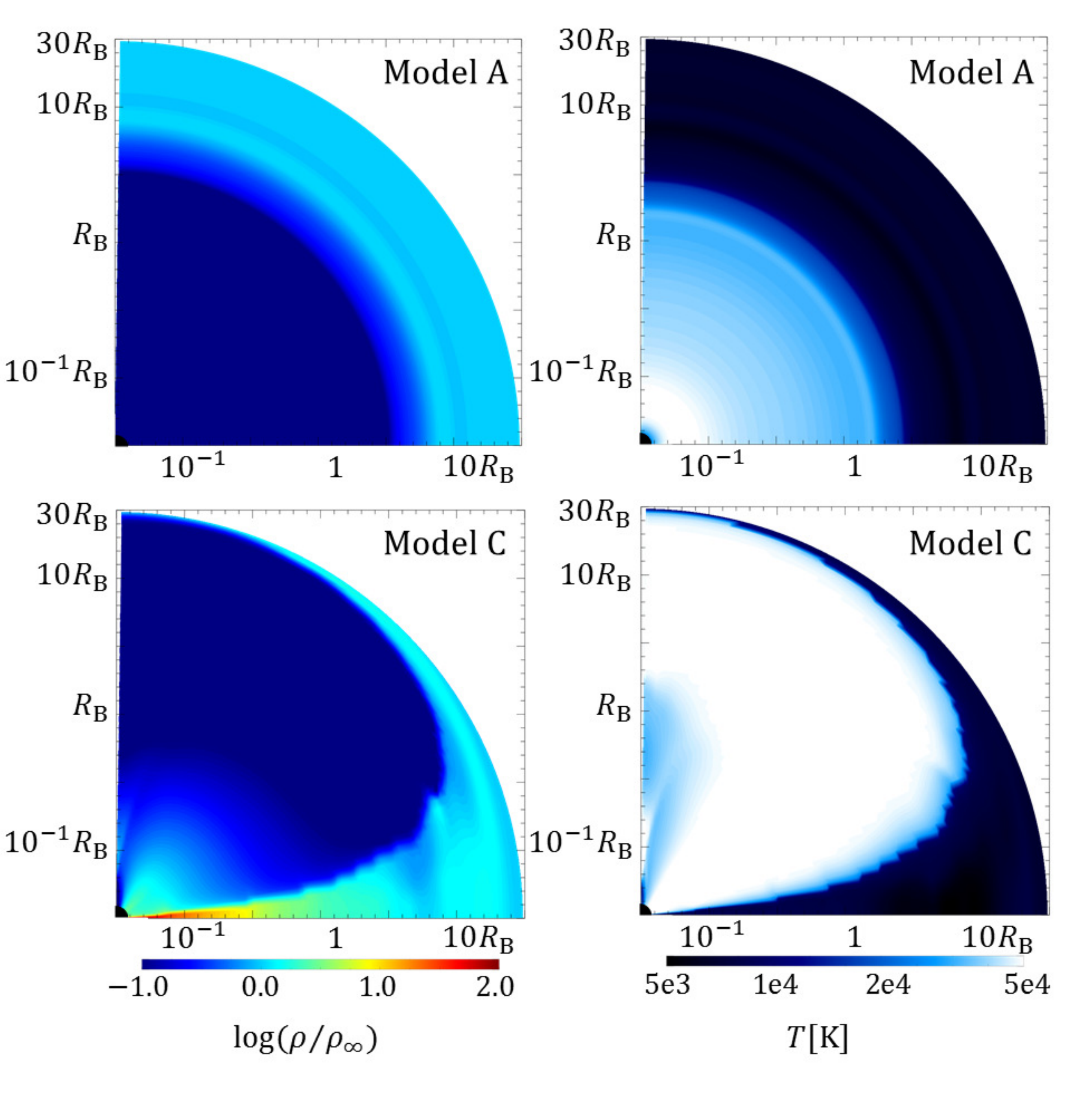}
\end{center}
\vspace{-4mm}
\caption{
Two-dimensional distribution of gas density (left) and temperature (right) 
for Model A (upper) and Model C (bottom), respectively, at 
$t=1.06 \times 10^5~{\rm yr}$ ($\approx 13~t_{\rm dyn}$). 
In Model C (anisotropic radiation), the ionized region does not exist around the equatorial plane,
from which a higher accretion rate is allowed.
}
\label{rho-temp-cont}
\end{figure*}

\subsection{Initial and boundary conditions}
\label{sec:IBC}

We set a computational domain of $r_{\rm min} \leq r \leq r_{\rm max}$ and $0 \leq \theta \leq \pi$. 
Here we adopt $r_{\rm min} = 0.035~R_{\rm B}$, $r_{\rm max} = 30~R_{\rm B}$ for Models A, B, and C,
and $r_{\rm min} = 0.01~R_{\rm B}$, $r_{\rm max} = 6~R_{\rm B}$ for Models D and E.
In Models A-C, the computational domain is located relatively outward 
because the size of the ionized region tends to larger than the Bondi radius.
Thus, we set logarithmically-spaced grids in the radial direction for Models A-C
to simulate the flow within the Bondi radius in high resolution.
We also employ power-law-spaced grids in Models D and E
to resolve the structure of outflows in the outer region ($r>R_{\rm B}$), 
as well as inflows in the inner region ($r<R_{\rm B}$).
%
We set uniformly-spaced grids in the polar direction. 
The number of the grid points is set to $(N_r, N_{\theta}, N_{\phi}) = (100,120,1)$.
Note that our simulations do not consider accretion flows within $r_{\rm min}$.
Instead, we assume properties of radiation emitted from the central region (see \S\ref{sec:BE})
and discuss gas accretion outside from the Bondi radii.

As our initial conditions, we set a neutral uniform and static (${\boldsymbol v}= 0$) gas cloud with the density 
$n_{\infty}=10^5~\cc$ and temperature $T_{\infty}=10^4~\K$.
The BH masses are assumed to be constant throughout our simulations.
We impose the absorption inner-boundary conditions for the gas density, 
gas pressure and velocity to be damped smoothly \citep[e.g.][]{kato+04}, 
and the free outer-boundary conditions for three components of the velocity 
and the specific entropy.
We also fix the same gas density at $r=r_{\rm max}$ as the initial value for grids with a inflow velocity 
i.e., $v_r(r=r_{\rm max}) < 0$, otherwise the free boundary conditions are imposed for the density.
The reflection symmetry with respect to the polar axis is imposed for 
non-radial components of the velocity.


\section{Results}

\subsection{Effects of anisotropic radiation}
\label{sec:aisotropy}

\begin{figure*}
\begin{center}
\includegraphics[width=13.5cm]{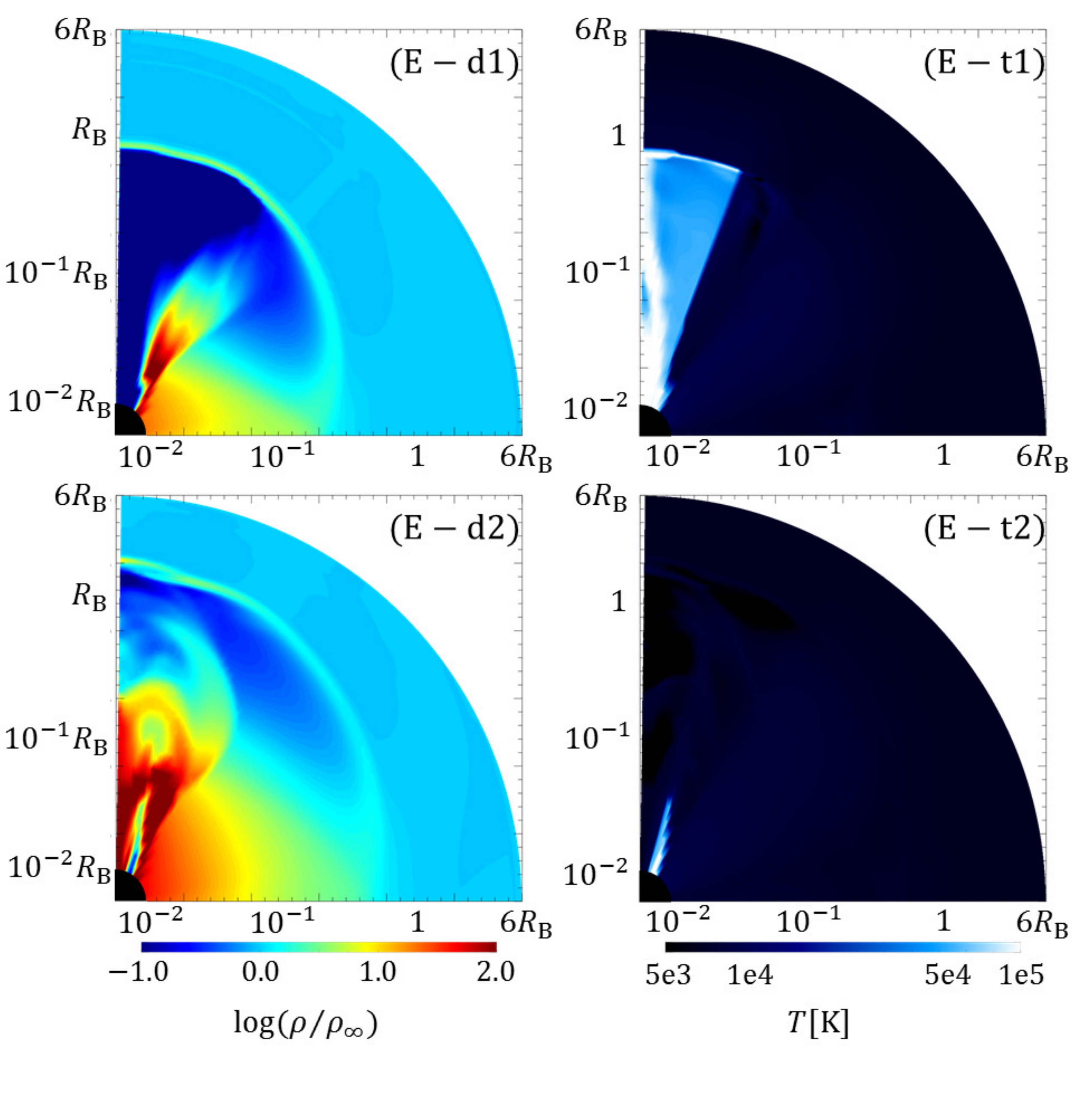}
\end{center}
\vspace{-4mm}
\caption{
Two-dimensional distribution of gas density (left) and temperature (right) 
in Model E ($M_{\rm BH}=5\times10^5~\msun$ and $N=4$)
in the early stage at $t=6.2\times10^5\ {\rm yr}~(\simeq 0.15~t_{\rm dyn})$ (top), 
and in the late stage (after the transition)
at $t=2.1\times10^6\ {\rm yr}~(\simeq 0.49~t_{\rm dyn})$ (bottom), 
respectively.
}
\label{rho_snapshots}
\end{figure*}

Figure \ref{time-lmd} presents the time evolution of gas accretion rates onto a BH with $M_{\rm BH}=10^3~\msun$
for Model A ($N=0$), B ($N=2$), and C ($N=4$).
For the isotropic radiation, the accretion occurs episodically and 
the time-averaged rate ($\sim 0.9~\dot{M}_{\rm Edd}$) is below the Eddington rate.
For the anisotropic radiation, the accretion is less episodic and their rates tend to increase with time
because of continuous accretion of neutral gas through the equatorial plane as shown later.
The time-averaged accretion rate becomes higher with $N$ (i.e., more anisotropic radiation).
In Model C, the accretion rate is as high as $\sim 44~\dot{M}_{\rm Edd}$ at the end of the simulation,
where the radiation luminosity is higher than the Eddington value, namely $\sim 3.5~L_{\rm Edd}$.
We continue the simulation until the ionization front reaches the outer boundary at 
$t \simeq  1.2\times 10^5$ yr, which is much longer than
$t_{\rm dyn} \equiv \pi\sqrt{R_{\rm B}^3/(8GM_{\rm BH})}
\simeq 8.4 \times 10^3~{M_{\rm BH,3}}~{\rm yr}$.
The accretion rate and the radiative luminosity at the end of each simulation
are summarized in Table 1.

In Figure \ref{rho-temp-cont}, we show two-dimensional distribution of the gas density and temperature
at $t=1.1 \times 10^5$ yr in Models A and C, respectively.
In the isotropic case (Model A), radiation from the central region heats up
and ionizes the surrounding gas equally in all directions. 
When the ionized region expands and the temperature outside the Bondi radius increases, 
gas supply from large radii is suppressed
since $\dot{M}_{\rm B} \propto T_\infty^{-3/2}$,
leading to episodic accretion (see Figure \ref{time-lmd}) as shown in previous studies 
\citep[e.g.][\citetalias{inayoshi+16}]{Ciotti_Ostriker_2001,Park_Ricotti_2011,Park_Ricotti_2012}.
Note that radiation heating is a dominant process for suppressing the gas accretion,
since the radiation luminosity is below the Eddington value.

\begin{figure*}
\begin{center}
\includegraphics[width=18cm]{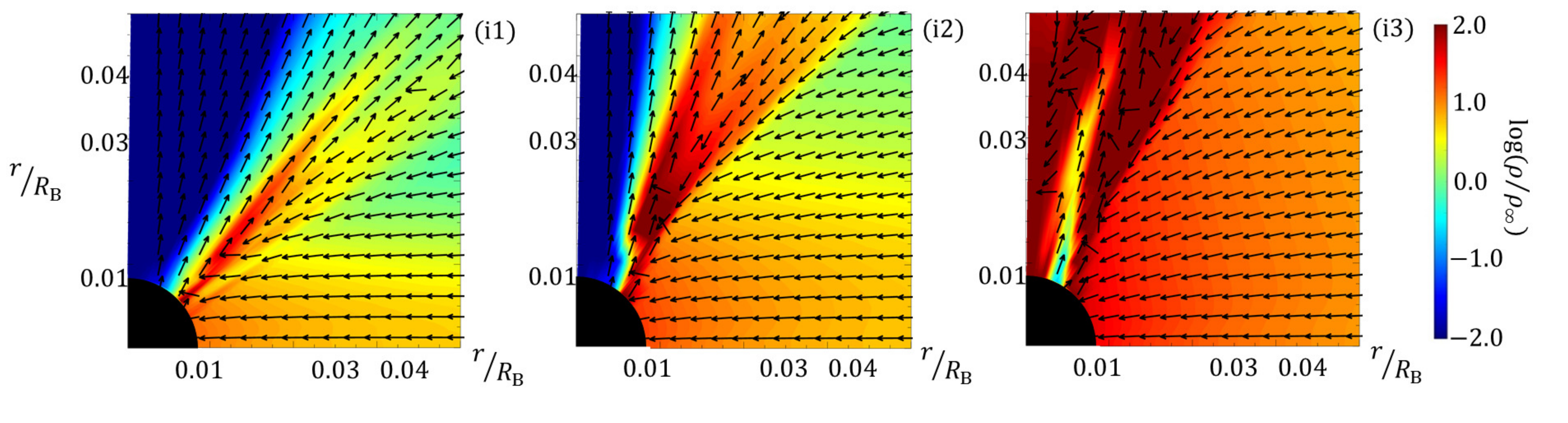}
\end{center}
\vspace{-2mm}
\caption{
Two-dimensional distribution of gas density and velocity structure in the central region 
($r_{\rm min} \leq r \la 0.05R_{\rm B}$) 
for Model E ($M_{\rm BH}=5\times 10^5~\msun$).
Each panel corresponds to the time at 
(i1) $t = 3.3 \times 10^5{\rm yr}~(\simeq 0.08~t_{\rm dyn})$,
(i2) $t = 6.2 \times 10^5{\rm yr}~(\simeq 0.15~t_{\rm dyn})$,
and 
(i3) $t = 2.1 \times 10^6{\rm yr}~(\simeq 0.49~t_{\rm dyn})$.
}
\label{tran_vecr}
\end{figure*}

In the anisotropic case (Model C), the shape of the ionized region is no longer isotropic. 
While the ionization front expands toward the polar directions,
the radiation flux in the equatorial direction ($\theta = \pi/2$) is not intense enough to form an ionized region.
Since the radiation luminosity viewed from the polar directions
is significantly higher than the Eddington luminosity, 
strong outflows are launched by the radiation force onto electrons.
The outflows collide with the ambient gas at $r\ga R_{\rm B}$ and form strong shocks 
with a high temperature of $\ga 5\times 10^4~\K$.
The gas near the equator remains neutral and the temperature is kept at $T\simeq 8000~\K$ 
due to efficient atomic hydrogen cooling.
Therefore, the gas can accrete onto the central BH through the neutral region 
increasing the density, unimpeded by radiation feedback.
The radial profiles at $\theta =\pi/2$ of the density and inflow velocity approach 
those for a Bondi solution with $T\sim 8000~\K$
($\rho \propto r^{-3/2}$ and $v_{\rm r}\propto r^{-1/2}$ at $r\la R_{\rm B}$).

The accretion rate through the equatorial plane is estimated
as $\dot{M}_{\rm HI} \simeq \dot{M}_{\rm B}(\Omega_{\rm HI} / 4\pi)$,
where $\Omega_{\rm HI}$ is the solid angle of the neutral region.
Since a half angle of the neutral region is $\Theta_{\rm HI} \equiv \pi/2-\theta \approx 5^{\circ}$ 
and $\dot{M}_{\rm B} \approx 670~\dot{M}_{\rm Edd}$, 
the inflow rate of the neutral gas is estimated as $\dot{M}_{\rm HI}\simeq 58~\dot{M}_{\rm Edd}$.
This value agrees to the simulation result of $\simeq  44~\dot{M}_{\rm Edd}$.
Unlike our setups, \cite{sugimura+16} assumed a shadow region with a half angle of 
$\Theta_{\rm Shad} = 10^{\circ}-45^{\circ}$, within which the radiation flux is completely shut off 
and accretion of neutral gas is allowed.
In this case, they also found that gas accretion rates are given by 
$\dot{M} \simeq \dot{M}_{\rm B}(\Omega_{\rm Shad} / 4\pi)$, which is consistent with our results.

\subsection{Transition to wholly neutral accretion}
\label{sec:tran}

Next, we examine a case with the highest BH mass of $5\times 10^5~\msun$ 
(Model E) but with the same radiation anisotropy ($N=4$) as in Model C.
Figure \ref{rho_snapshots} shows two-dimensional distribution of the gas density 
(left panels) and temperature (right panels)
in the early stage at $t\simeq 0.15~t_{\rm dyn}$ (top panels) and 
the late stage at $t\simeq 0.49~t_{\rm dyn}$ (bottom panels), respectively.
Initially, the ionized region expands to $r\sim R_{\rm B}$ preferentially toward the polar directions,
and gas accretion
is allowed through the neutral region
(top panels in Figure \ref{rho_snapshots}).
Unlike in Model C with lower $M_{\rm BH}$, however,
neutral and dense gas inflows through the equator totally cover the central region.
Since ionizing photons from the center are absorbed by the surrounding neutral gas,
the radiation flux in all directions (even in the polar directions) is blocked.
As a result, the ionized region is confined within $r \sim 0.04R_{\rm B}$ due to radiative recombination
(bottom panels in Figure \ref{rho_snapshots}).

\begin{figure}
\begin{center}
\vspace{-5mm}
\includegraphics[width=6cm, angle=270]{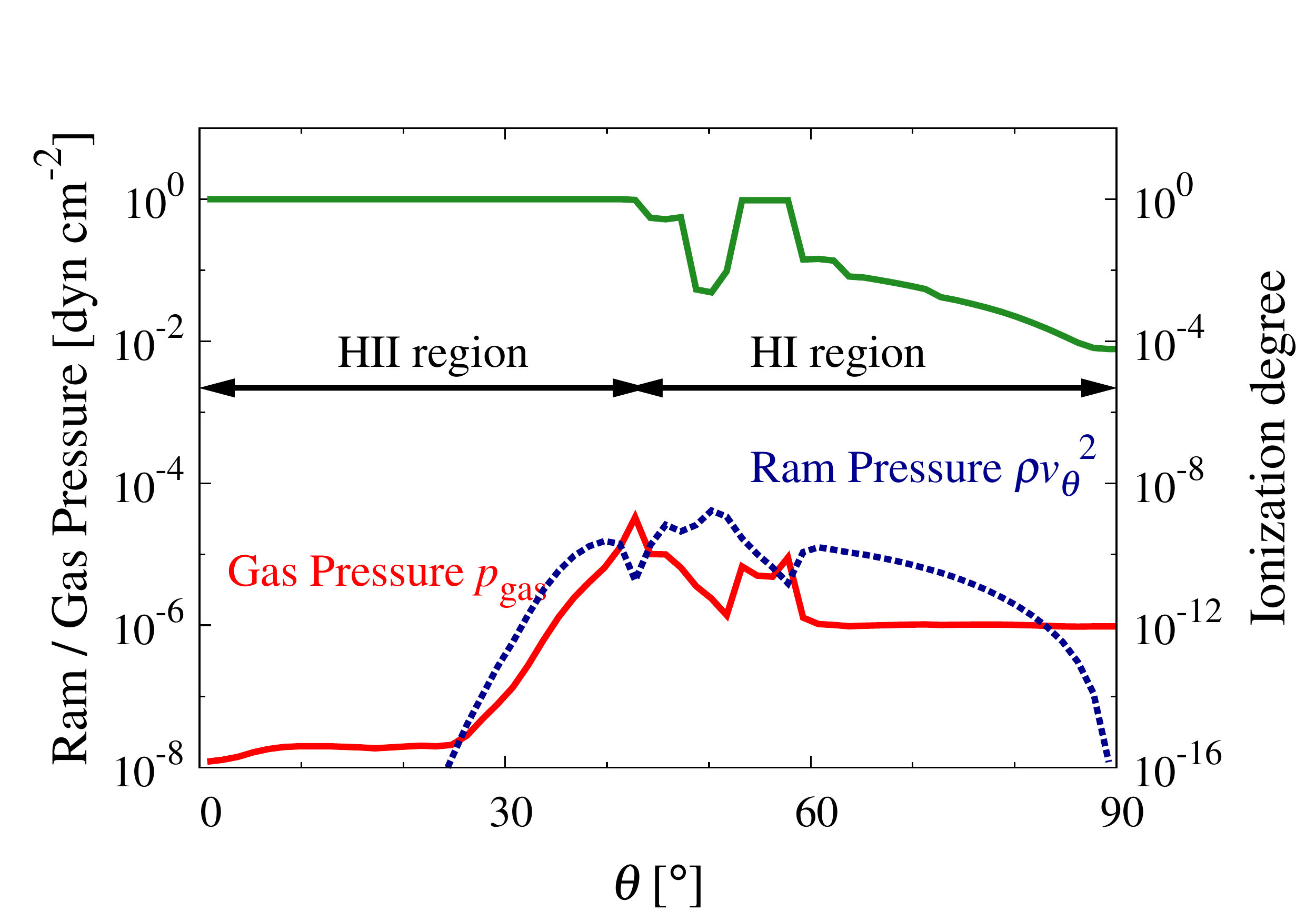}
\caption{ 
Profiles of ram pressure $\rho v_{\theta}^2$ (blue), 
gas pressure $p_{\rm gas}$ (red),
and the ionization degree (green)
in the northern hemisphere from the polar ($\theta =0^\circ$) to the equator ($\theta =90^\circ$)
at $r=0.02R_{\rm B}$ in Model E.
The elapsed time is the same as that of panel (i1) in Figure \ref{tran_vecr}. 
}
\label{tran_ramp}
\end{center}
\end{figure}

In order to understand how and why this transition of the accretion episode occurs,
we examine structure of the gas flow 
in the central region ($r_{\rm min} \leq r \la 0.05R_{\rm B}$).
Figure \ref{tran_vecr} shows the density distribution and the velocity vectors of the flows.
Each panel shows the time sequence from the left (before the transition) to the right (after the transition).
Figure \ref{tran_ramp} also presents profiles of the ionization degree, 
the ram pressure ($\rho v_{\theta}^2$) and the gas pressure ($p_{\rm gas}$)
as functions of the polar angle at $r=0.02R_{\rm B}$. 
Before the transition shown in Figure \ref{tran_vecr} (panel i1),
a geometrically thin and dense shock layer forms 
along the direction of $\theta \approx 45^{\circ}$,
where the boundary between ionized and neutral gas is located (see Figure \ref{tran_ramp}).
The inflow from the neutral region crosses the shock with a supersonic velocity, 
namely $|v_{\theta}|>c_{\rm s}$ and is decelerated
($c_{\rm s}$ is the local sound speed).
The shocked gas is ionized and accelerated outward in radial direction by strong radiation force.
As shown in Figure \ref{tran_ramp}, ram pressure of the neutral gas inflow slightly dominates 
gas pressure of the ionized region before the transition.
As a result, the neutral gas inflow pushes the shock layer from the equator to the polars.
Panel (i2) in Figure \ref{tran_vecr} shows that 
a dense clump forms at the shock front by compression caused by
continuous accretion of neutral gas and strong radiation force from the center.
Although some clumps are ejected by radiation force, such clump formation occurs more 
frequently in the late stage.
Eventually, the central radiating region is totally covered by dense clumps of neutral gas.

\begin{figure}
\begin{center}
\includegraphics[width=8.2cm]{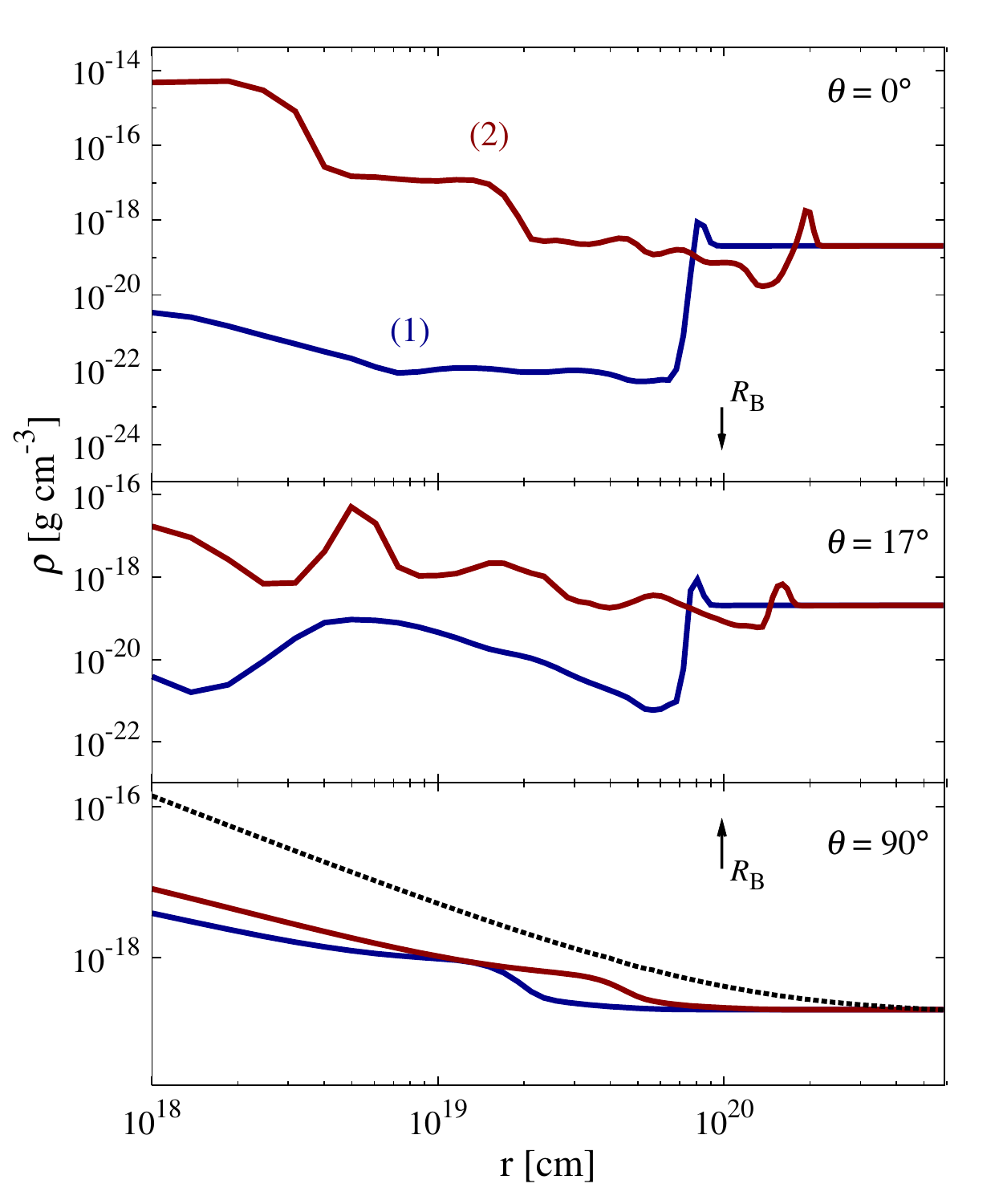}
\end{center}
\vspace{-3mm}
\caption{Radial profiles of the gas density in Model E
along three different directions of $\theta=0^\circ$ (polar), $\theta =17^\circ$,
and $\theta=90^\circ$ (equator)
in the early stage at $ t = 0.15~t_{\rm dyn}$ (blue) and 
in the late stage at $t = 0.49~t_{\rm dyn}$ (red).
The black curve represents an isothermal Bondi profile with $T=10^4 {\rm K}$.
\label{rho_temp_pro}}
\end{figure}

\begin{figure}
\vspace{-3mm}
\begin{center}
\includegraphics[width=8.2cm]{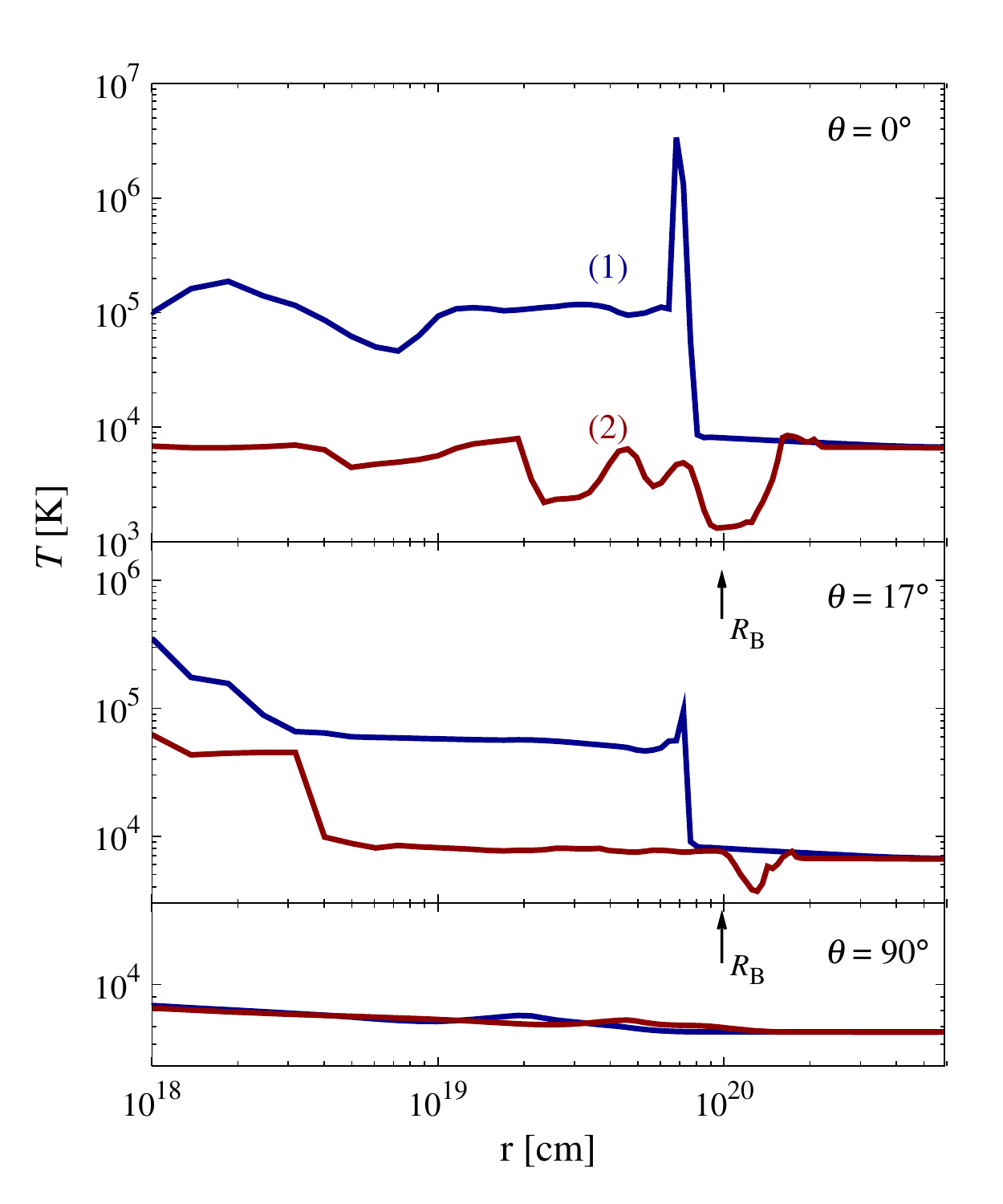}
\end{center}
\vspace{-5mm}
\caption{
Same as Figure \ref{rho_temp_pro}
but for the gas temperature.}
\label{tempr}
\end{figure}

\begin{figure}
\vspace{-6mm}
\begin{center}
\includegraphics[width=8.2cm]{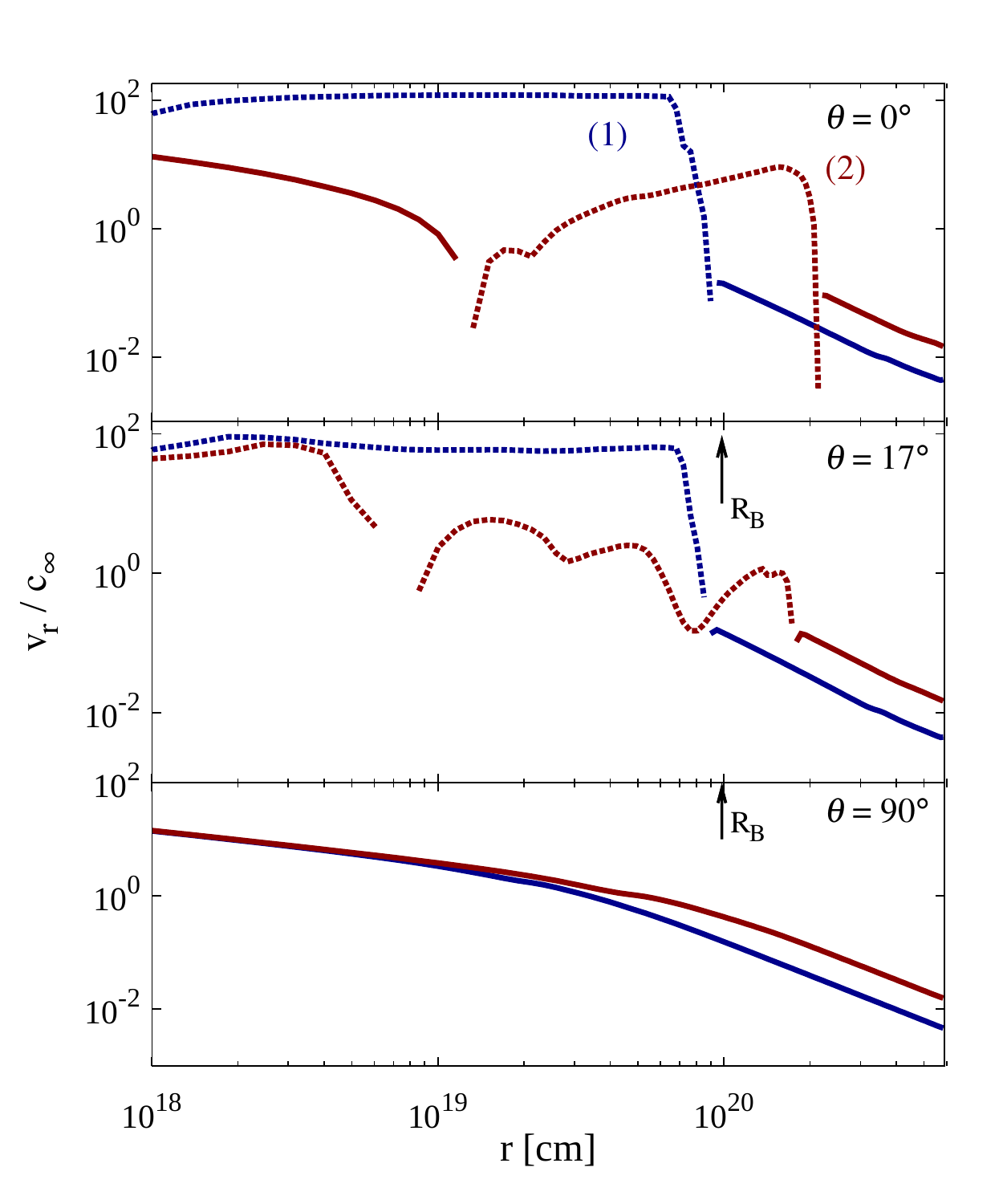}
\end{center}
\vspace{-5mm}
\caption{
Same as Figure \ref{rho_temp_pro}
but for the radial velocity.
Solid and dashed curves present the inflow and outflow velocity, respectively.
\label{vrr}}
\end{figure}

Figures \ref{rho_temp_pro}, \ref{tempr} and \ref{vrr} present radial profiles of the gas density, 
temperature, and radial velocity
along three different directions: $\theta=0^\circ$ (polar), $17^\circ$,
and $90^\circ$ (equator) in the early stage at $ t = 0.15~t_{\rm dyn}$ (blue) and 
in the late stage at $t = 0.49~t_{\rm dyn}$ (red). 
The black dashed curve represents the isothermal Bondi profile with $T=10^4 {\rm K}$.
In the early stage before the transition, radiation force and heating affect properties 
of the accreting material.
Inside the ionized region, the gas is depleted and a density cavity is created ($\theta =0^\circ$ and $17^\circ$).
Near the edge of the ionized region, colliding outflows due to the radiation force on electrons (dashed curves in Figure \ref{vrr}) 
produce strong shocks with a high temperature of $\sim 10^6~\K$.
On the other hand, neutral gas accretes through the equator keeping the temperature at $T\simeq 8000~\K$,
as shown in \S\ref{sec:aisotropy} (Model C).
After the transition, the ionized gas becomes neutral due to radiative recombination and the density cavity is filled.
The gas temperature settles down to $\sim 10^4$ K in all directions because of radiative cooling by atomic hydrogen.
Along the direction of $\theta=17^{\circ}$, the ionized region still exists but its size is too small ($\sim 0.04R_{\rm B}$) 
to affect gas dynamics at the Bondi radius.
As a result of absorption of anisotropic radiation by neutral hydrogen, 
the radiation momentum produces warm outflows with $T \simeq  8000$ K toward 
narrow conical regions along the polars ($\theta \sim 0^\circ-20^\circ$).


\begin{figure}
\vspace{-2.2mm}
\hspace{-0.7mm}
\includegraphics[width=8.6cm]{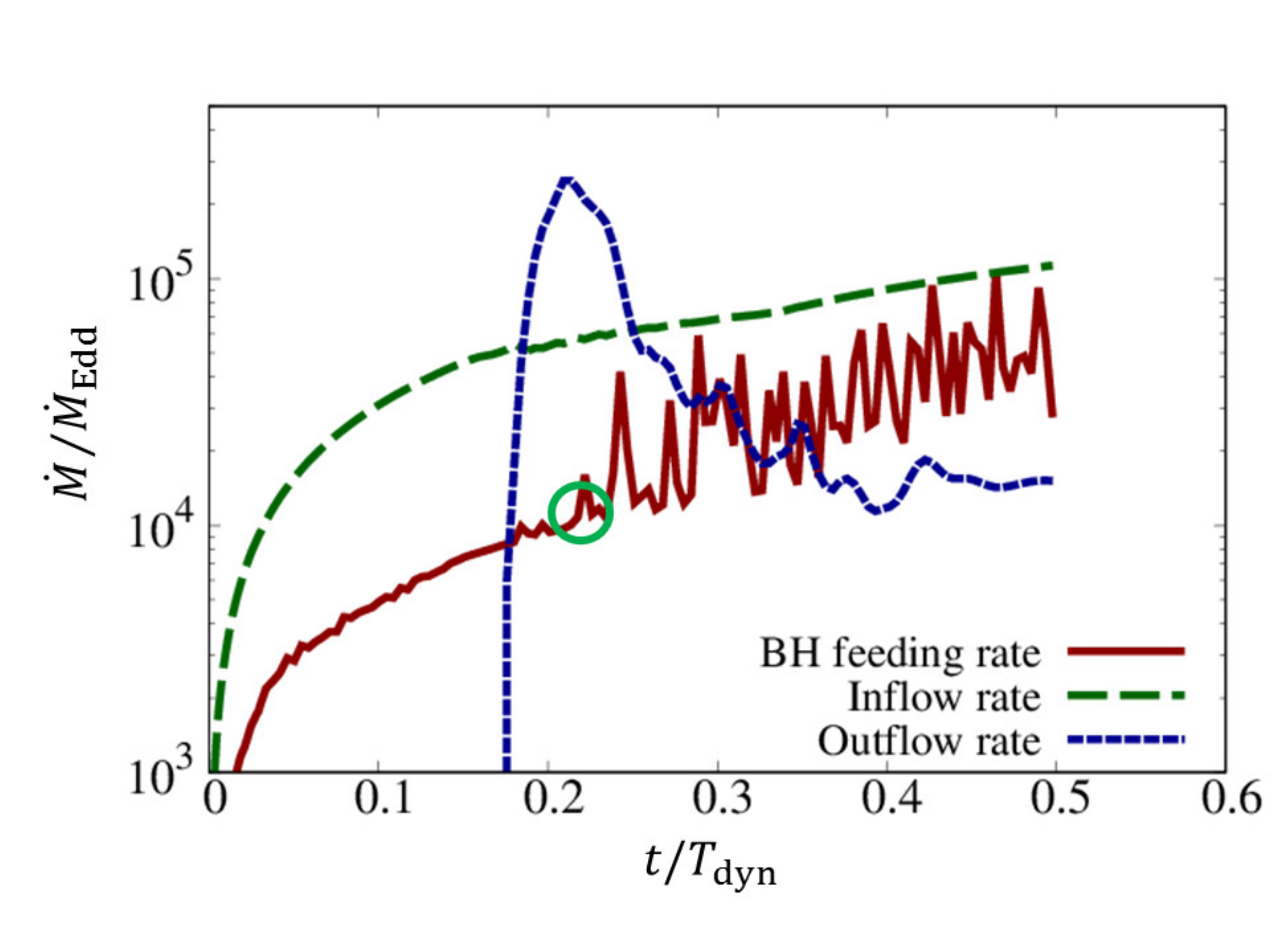}
\caption{
Time evolution of inflow (green long-dashed) and outflow (blue short-dashed) rates at the Bondi radius 
and the BH feeding rate (red solid) in Model E ($M_{\rm BH}=5\times 10^5~\msun$).}
\label{t_mdinout}
\end{figure}

Figure \ref{t_mdinout} shows the time evolution of the inflow (green long-dashed) and outflow (blue short-dashed) rate 
measured at the Bondi radius, and the BH feeding rate (red solid).
The normalized BH feeding rate of $\dot{M}/\dot{M}_{\rm Edd}$ continuously increases with 
the inflow rate from the Bondi radius.
An open circle indicates the epoch of the transition to the wholly neutral accretion phase.
After the transition, the feeding rate becomes highly episodic because strong radiation force and 
high ram-pressure are tightly balanced at the innermost region.
At the end of the simulation, the central BH is fed through the equator at the time-averaged rate 
of $\sim 5 \times 10^4~L_{\rm Edd}/c^2$, which corresponds to $\sim 50~\%$ of the inflow rate 
from the Bondi radius.
On the other hand, the accreting material is accelerated outward into the bipolar directions
($\theta \la 20^\circ$) due to absorption of the radiation momentum with $\gg L_{\rm Edd}/c$.
The outflow rate ends up $\simeq 1.5\times 10^4~\dot{M}_{\rm Edd}$, 
which is $\sim 30~\%$ of the BH feeding rate.
Since the Mach number of the outflowing matter is as high as five at the Bondi radius (see Figure \ref{vrr}),
the typical velocity is estimated as $\simeq 50~{\rm km~s}^{-1}$.
\cite{sugimura+16} reported that hot and ionized outflows with $T\simeq 7\times 10^4~{\rm K}$
are launched from a rapid accreting system exposed to strong anisotropic radiation,
which is consistent with our cases where the transition does not occur as in Model C.
On the other hand, our simulation in Model E suggests that warm and neutral outflows with 
$T\simeq 8000~{\rm K}$ are likely to be produced because of the transition to the wholly neutral phase.

\begin{figure}
\vspace{-3mm}
\includegraphics[width=8.7cm]{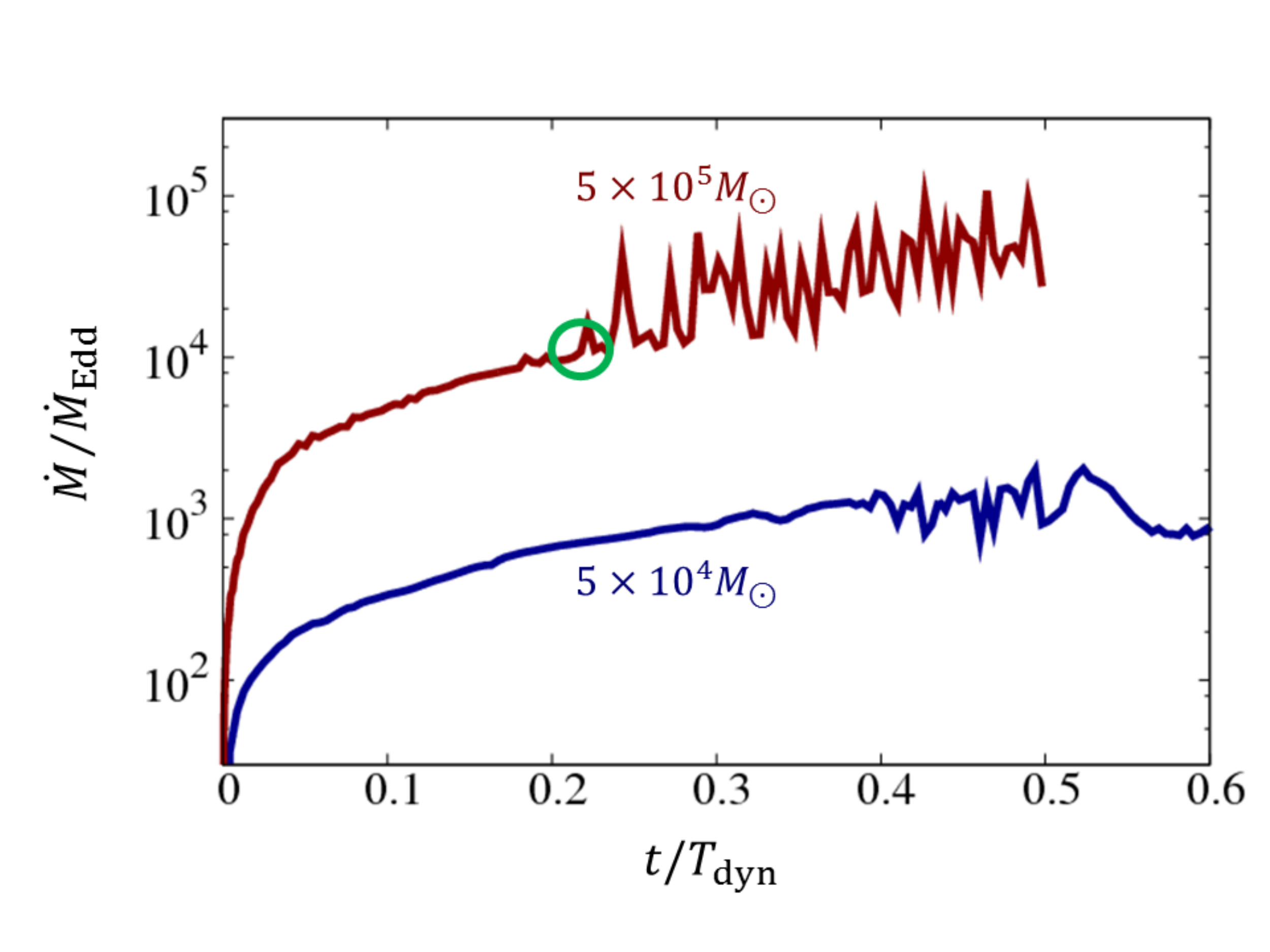}
\vspace{-8mm}
\caption{
Time evolution of the accretion rates for two cases 
with $M_{\rm BH}=5\times 10^4~\msun$ (Model D) and 
$M_{\rm BH}=5\times 10^5~\msun$ (Model E).
An open circle presents the epoch where the ionization region 
collapses by radiative recombination.
The transition occurs for $M_{\rm BH} \ga 5 \times 10^5~\msun$.
}
\label{t_md_massive}
\end{figure}

In Figure \ref{t_md_massive}, we show the BH feeding rates for two cases with 
$M_{\rm BH} = 5 \times 10^4~\msun$ (blue, Model D) and $5\times 10^5~\msun$ (red, Model E) for comparison. 
The normalized accretion rate in Model D is saturated at a constant value unlike in Model E. 
In fact, we do not find a transition to a wholly neutral phase within the entire simulation time.
Even without the transition, neutral gas inflows have non-radial motions and dense clumps are formed 
at the boundary between the ionized and neutral region as in Model E.
However, radiation force from the center dominates ram-pressure of the infalling clumps,
and thus such clumps do not cover the central radiation source.
Therefore, this result implies the critical BH mass for the transition is located around 
$M_{\rm BH} \sim 5 \times 10^5~\msun$ for the initial conditions we adopted
($n_\infty=10^5~\cc$ and $T_\infty=10^4~\K$).

\section{Summary and discussion}
\label{sec:discuss}

In this paper, we study rapid growth of BHs with a mass range of $10^3\leq M_{\rm BH} \leq 5\times 10^5~\msun$
via gas accretion exposed to anisotropic super-Eddington radiation.
Radiation flux from an accretion disk around the nuclear BH is likely to be collimated 
toward the bipolar directions.
To address the effects of the anisotropic radiation on the accretion flow, 
we perform two-dimensional radiation hydrodynamical simulations, 
including multifrequency radiation transfer and non-equilibrium primordial chemistry.
We here consider a  BH embedded in an initially uniform gas cloud 
with the density of $n_\infty = 10^5~\cc$ and temperature of $T_\infty=10^4~\K$.
We have found that hyper-Eddington accretion from the Bondi radius can occur 
when the radiation is more anisotropic and the BH mass is higher. 
Moreover, when the BH mass is as large as $M_{\rm BH}\ga 5\times 10^5~\msun$,
ionized regions around the BH collapse due to super-sonic non-radial motions of 
neutral gas inflows and efficient radiative recombination.
Due to absorption of the radiation momentum,
the neutral gas is accelerated outward to narrow conical regions along the polars,
producing warm outflows with $T \simeq 8000~\K$.

We adopt a simple model for anisotropic radiation
from the central accretion disk, namely, $F_{\rm rad}(\theta)\propto \cos^N \theta$. 
As a result of the anisotropic radiation, rapid accretion is allowed from the equatorial plane. 
This assumption would be reasonable at the vicinity of the central BH 
because a super-Eddington accretion disk is optically and geometrically thick,
so that the radiation flux is mildly collimated to the polars
\citep[e.g.,][]{ohsuga+05,Jiang_2014,Sadowski_2015}.
However, while the radiation propagates outward,
such anisotropic radiation fields could be smoothed out 
because of radiative diffusion.
In order to explore the anisotropy of the emergent radiation, therefore
radiation-hydrodynamical simulations with a wide dynamical range are required.

Recently, \cite{sugimura+16} have performed two-dimensional simulations 
of accreting BHs under anisotropic radiation and found that anisotropic radiation 
allows gas accretion through the equatorial plane.
Although our calculations have been pursued independently from their work, 
our simulation setups are similar, mainly except two things.
First, they assumed the existence of a shadow region caused by 
attenuation due to a geometrically thick disk and/or disk winds 
well inside their computation domains.
Second, our simulations resolve even smaller scales, namely $\sim 1\%$ of the Bondi radius.
We can stress that our setup is more preferable to follow non-radial motions of 
dense and neutral inflows, which lead to a transition to a wholly neutral accretion phase.
Conversely,  \cite{sugimura+16} did not find such transitions, 
probably because (1) the BH mass adopted in their simulations is lower than
that in our Model E and (2) the radius of the inner-most grid in their simulations 
is larger than that in our simulations.

We discuss the subsequent evolution of the accretion flows in Model E 
($M_{\rm BH}=10^{5.7}~\msun$).
The gas profile approaches an isothermal Bondi solution with $T\sim 10^4$~K,
where $\dot{M}_{\rm B}/\dot{M}_{\rm Edd}\simeq 3 \times 10^5~n_{\infty.5}M_{\rm BH,5.7}$
if the ram pressure of the neutral inflow dominates the radiation force.
This condition is written as
$L/L_{\rm Edd} \la 116~M_{\rm BH,5.7}^{3/2}n_{\infty,5} (r_{\star}/10^{18}~{\rm cm})^{-1/2}$
\citep{Sakurai_2016}, where $r_\star$ is the location of the photosphere, from which the radiation emerges actually.
Note that the photosphere is located inside the innermost grid ($r_\star <r_{\rm min}=10^{18}~{\rm cm})$.
Assuming that the BH is fed at the Bondi rate, the radiation luminosity toward the polar directions
is estimated as $L/L_{\rm Edd}
\approx 106$ (see Eq.\ref{frad_in}).
Therefore, we expect that the overall behaviors of the accretion flows in Model E would be maintained.

As shown in Figure \ref{tran_vecr}, non-radial motions of neutral inflows 
near the inner boundary are essential for leading to the transition to a wholly neutral phase.
However, our simulations neglect non-radial components in radiation flux $F_{\rm rad}^{\theta}$,
which can be produced by radiative recombination in the ionized region\footnote{
The non-radial components in radiation flux is unlikely to be produced
by electron scattering because the ionized gas is optically thin even near the innermost region.}.
The radiation flux due to recombination is 
estimated as $F_{\rm rad}^{\theta} \sim 4\pi \eta r_{\rm min}$,
where $\eta \sim \langle h\nu_{\rm rec} \rangle \alpha_{\rm B}n_{\rm e}^2
\sim 10^{-18}(n_{\rm e}/10^3~\cc)^2~{\rm erg\ cm^{-3} s^{-1}}$ is the emissivity,
$ \langle h\nu_{\rm rec} \rangle$ is the average energy of recombination photons,
$\alpha_{\rm B}$ is the case B recombination rate coefficient,
and $n_{\rm e}$ is the numb density of electrons.
Thus the non-radial radiation flux is estimated as
$|F_{\rm rad}^{\theta}/F_{\rm rad}^r| \sim 10^{-8}$ at the innermost region, and
the radiation force in non-radial directions is negligible.

Throughout this paper, we neglect formation of molecular hydrogen (H$_2$),
which is the main coolant in primordial gas below $\sim 10^4~\K$.
We briefly mention that H$_2$ cooling is unlikely to affect the properties of warm/neutral outflows.
In a warm and dense primordial gas cloud ($n\ga 10^5~\cc$ and $T\sim 8000~\K$),
the H$_2$ fraction relative to atoms is determined by the balance between formation through electron-catalyzed 
reactions and collisional dissociation with gas particles \citep{omukai_2001}.
The resultant fraction is estimated as $\sim 10^{-8}$ as long as the temperature is kept at $>4000~\K$, 
below which the collisional dissociation becomes inefficient \citep{IOT14}.
As shown in Figure \ref{tempr}, the temperature cools down to $\simeq 1000-2000~\K$
due to adiabatic expansion at $r>2\times 10^{19}$ cm in the outflowing matter ($\theta =0^\circ$).
In this lower-temperature region, far-ultraviolet radiation in the Lyman-Werner (LW) bands 
($h\nu_{\rm LW}=10.2-13.6$ eV) from the central region plays an important role to determine the H$_2$ fraction.
We can estimate the LW flux in the optically thin limit as 
\begin{align}
F_{\rm LW,thin}& = f_{\rm LW}(N+1)\frac{2\left[1+\ln(\dot{m}/20)\right]~L_{\rm Edd}}
{4\pi r^2\langle \nu_{\rm LW}\rangle}\nonumber\\
&\simeq  1.2\times 10^{-11}~{\rm erg~s^{-1}~cm^{-2}~Hz^{-1}}\\
&\times
\left(\frac{f_{\rm LW}}{0.3}\right)
\left(\frac{N+1}{5}\right)
\left(\frac{M_{\rm BH}}{5\times 10^5~\msun}\right)^{-1}
\left(\frac{r}{R_{\rm B}}\right)^{-2}, \nonumber
\end{align}
where $f_{\rm LW}$ is the ratio of the LW flux to the bolometric value and 
$\langle \nu_{\rm LW}\rangle = 12~{\rm eV}/h~(=2.9\times 10^{15}~{\rm Hz})$.
Since the H$_2$ column density is $N_{\rm H_2}\sim 10^{17}~{\rm cm}^{-2}(n_\infty/10^5~\cc)(r/R_{\rm B})$ at most, 
the emergent LW flux can be reduced by the H$_2$ self-shielding effect.
The shielding factor is simply estimated as 
$f_{\rm shield}\simeq (N_{\rm H_2}/10^{14}~{\rm cm}^{-2})^{-3/4}$ for a higher column density 
(\citealt{draine_1996}, see also \citealt{WHB_2011})
and thus $F_{\rm LW}=f_{\rm shield}F_{\rm LW,thin}\simeq 7\times 10^{-14}
~{\rm erg~s^{-1}~cm^{-2}~Hz^{-1}}$, which is 
higher than the critical LW flux\footnote{The critical LW flux depends on the radiation spectrum, namely
on the ratio of the flux at $2$ eV to the LW flux (e.g., Sugimura, Omukai \& Inoue 2014). 
This is because photons with $\sim 2$ eV dissociate H$^-$ ions, 
which is an intermediate product in the H$_2$ formation channels 
(${\rm H} + {\rm e}^- \rightarrow {\rm H}^- + \gamma$; ${\rm H}^- + {\rm H} \rightarrow {\rm H}_2 + {\rm e}^-$)
and contribute to H$_2$ dissociation indirectly \citep{omukai_2001}.}, 
$F_{\rm LW,crit}\sim 10^{-19}-10^{-16}$ erg s$^{-1}$ cm$^{-2}$ Hz$^{-1}$
\citep[e.g.][]{sugimura_2014,IT_2015,Latif_2016,WHB_2017}
required to dissociate H$_2$ and suppress H$_2$ cooling in the warm and neutral gas
even at $r\sim 10^{20}$ cm.


\section*{Acknowledgements}
We thank Zolt\'an Haiman and Kazuyuki Sugimura for useful discussions.
This work is partially supported by the Simons Foundation through the Simons Society of Fellows (KI)
and by JSPS Grant-in-Aid for Scientific Research (C) (15K05036 K.O.) and for Young Scientists (17K14260 H.R.T). 
This research was also supported by MEXT as ''Priority Issue on Post-K computer''
(Elucidation of the Fundamental Laws and Evolution of the Universe) and JICFuS.
Numerical computations were carried out on Cray XC30 at Center for Computational Astrophysics, 
National Astronomical Observatory of Japan.

{\small
\bibliography{ref.bib}

\begin{thebibliography}{64}
\expandafter\ifx\csname natexlab\endcsname\relax\def\natexlab#1{#1}\fi

\bibitem[{{Abel} {et~al}\mbox{.}(1997){Abel}, {Anninos}, {Zhang}, \&
  {Norman}}]{abel+97}
{Abel} T., {Anninos} P., {Zhang} Y., {Norman} M.~L., 1997, New Astron., 2, 181

\bibitem[{{Alexander} \& {Natarajan}(2014)}]{Alexander_2014}
{Alexander} T., {Natarajan} P., 2014, Science, 345, 1330

\bibitem[{{Alvarez}, {Wise} \& {Abel}(2009){Alvarez}, {Wise}, \&
  {Abel}}]{Alvarez_2009}
{Alvarez} M.~A., {Wise} J.~H., {Abel} T., 2009, \apjl, 701, L133

\bibitem[{{Anninos} {et~al}\mbox{.}(1997){Anninos}, {Zhang}, {Abel}, \&
  {Norman}}]{anninos+97}
{Anninos} P., {Zhang} Y., {Abel} T., {Norman} M.~L., 1997, New Astron., 2, 209

\bibitem[{{Becerra} {et~al}\mbox{.}(2015){Becerra}, {Greif}, {Springel}, \&
  {Hernquist}}]{2015MNRAS.446.2380B}
{Becerra} F., {Greif} T.~H., {Springel} V., {Hernquist} L.~E., 2015, \mnras,
  446, 2380

\bibitem[{{Begelman}(1979)}]{Begelman_1979}
{Begelman} M.~C., 1979, \mnras, 187, 237

\bibitem[{{Bondi}(1952)}]{Bondi_1952}
{Bondi} H., 1952, \mnras, 112, 195

\bibitem[{{Chon} {et~al}\mbox{.}(2016){Chon}, {Hirano}, {Hosokawa}, \&
  {Yoshida}}]{Chon_2016}
{Chon} S., {Hirano} S., {Hosokawa} T., {Yoshida} N., 2016, \apj, 832, 134

\bibitem[{{Ciotti} \& {Ostriker}(2001)}]{Ciotti_Ostriker_2001}
{Ciotti} L., {Ostriker} J.~P., 2001, \apj, 551, 131

\bibitem[{{Ciotti}, {Ostriker} \& {Proga}(2009){Ciotti}, {Ostriker}, \&
  {Proga}}]{2009ApJ...699...89C}
{Ciotti} L., {Ostriker} J.~P., {Proga} D., 2009, \apj, 699, 89

\bibitem[{{Devecchi} \& {Volonteri}(2009)}]{2009ApJ...694..302D}
{Devecchi} B., {Volonteri} M., 2009, \apj, 694, 302

\bibitem[{{Draine} \& {Bertoldi}(1996)}]{draine_1996}
{Draine} B.~T., {Bertoldi} F., 1996, \apj, 468, 269

\bibitem[{{Fan} {et~al}\mbox{.}(2004){Fan}, {Hennawi}, {Richards}, {Strauss},
  {Schneider}, {Donley}, {Young}, {Annis}, {Lin}, {Lampeitl}, {Lupton}, {Gunn},
  {Knapp}, {Brandt}, {Anderson}, {Bahcall}, {Brinkmann}, {Brunner}, {Fukugita},
  {Szalay}, {Szokoly}, \& {York}}]{Fan_2004}
{Fan} X. {et~al.}, 2004, \aj, 128, 515

\bibitem[{{Glover} \& {Jappsen}(2007)}]{glover+07}
{Glover} S.~C.~O., {Jappsen} A.-K., 2007, \apj, 666, 1

\bibitem[{{Haiman} \& {Loeb}(2001)}]{HaimanLoeb01}
{Haiman} Z., {Loeb} A., 2001, \apj, 552, 459

\bibitem[{{Inayoshi}, {Haiman} \& {Ostriker}(2016){Inayoshi}, {Haiman}, \&
  {Ostriker}}]{inayoshi+16}
{Inayoshi} K., {Haiman} Z., {Ostriker} J.~P., 2016, \mnras, 459, 3738

\bibitem[{{Inayoshi}, {Omukai} \& {Tasker}(2014){Inayoshi}, {Omukai}, \&
  {Tasker}}]{IOT14}
{Inayoshi} K., {Omukai} K., {Tasker} E., 2014, \mnras, 445, L109

\bibitem[{{Inayoshi} \& {Tanaka}(2015)}]{IT_2015}
{Inayoshi} K., {Tanaka} T.~L., 2015, \mnras, 450, 4350

\bibitem[{{Jiang}, {Stone} \& {Davis}(2014){Jiang}, {Stone}, \&
  {Davis}}]{Jiang_2014}
{Jiang} Y.-F., {Stone} J.~M., {Davis} S.~W., 2014, \apj, 796, 106

\bibitem[{{Kato}, {Mineshige} \& {Shibata}(2004){Kato}, {Mineshige}, \&
  {Shibata}}]{kato+04}
{Kato} Y., {Mineshige} S., {Shibata} K., 2004, \apj, 605, 307

\bibitem[{{Katz}, {Sijacki} \& {Haehnelt}(2015){Katz}, {Sijacki}, \&
  {Haehnelt}}]{2015MNRAS.451.2352K}
{Katz} H., {Sijacki} D., {Haehnelt} M.~G., 2015, \mnras, 451, 2352

\bibitem[{{King}(2003)}]{King_2003}
{King} A., 2003, \apjl, 596, L27

\bibitem[{{Kormendy} \& {Ho}(2013)}]{Kormendy_Ho_2013}
{Kormendy} J., {Ho} L.~C., 2013, \araa, 51, 511

\bibitem[{{Latif}, {Schleicher} \& {Hartwig}(2016){Latif}, {Schleicher}, \&
  {Hartwig}}]{Latif_2016}
{Latif} M.~A., {Schleicher} D.~R.~G., {Hartwig} T., 2016, \mnras, 458, 233

\bibitem[{{Lupi} {et~al}\mbox{.}(2016){Lupi}, {Haardt}, {Dotti}, {Fiacconi},
  {Mayer}, \& {Madau}}]{Lupi_2016}
{Lupi} A., {Haardt} F., {Dotti} M., {Fiacconi} D., {Mayer} L., {Madau} P.,
  2016, \mnras, 456, 2993

\bibitem[{{Madau}, {Haardt} \& {Dotti}(2014){Madau}, {Haardt}, \&
  {Dotti}}]{Madau_2014}
{Madau} P., {Haardt} F., {Dotti} M., 2014, \apjl, 784, L38

\bibitem[{{Madau} \& {Rees}(2001)}]{MadauRees01}
{Madau} P., {Rees} M.~J., 2001, \apjl, 551, L27

\bibitem[{{Milosavljevi{\'c}} {et~al}\mbox{.}(2009){Milosavljevi{\'c}},
  {Bromm}, {Couch}, \& {Oh}}]{Milosavljevic_2009}
{Milosavljevi{\'c}} M., {Bromm} V., {Couch} S.~M., {Oh} S.~P., 2009, \apj, 698,
  766

\bibitem[{{Mortlock} {et~al}\mbox{.}(2011){Mortlock}, {Warren}, {Venemans},
  {Patel}, {Hewett}, {McMahon}, {Simpson}, {Theuns}, {Gonz{\'a}les-Solares},
  {Adamson}, {Dye}, {Hambly}, {Hirst}, {Irwin}, {Kuiper}, {Lawrence}, \&
  {R{\"o}ttgering}}]{Mortlock_2011}
{Mortlock} D.~J. {et~al.}, 2011, \nat, 474, 616

\bibitem[{{Murray}, {Quataert} \& {Thompson}(2005){Murray}, {Quataert}, \&
  {Thompson}}]{Murray_2005}
{Murray} N., {Quataert} E., {Thompson} T.~A., 2005, \apj, 618, 569

\bibitem[{{Ohsuga} \& {Mineshige}(2007)}]{ohsuga_mineshige_2007}
{Ohsuga} K., {Mineshige} S., 2007, \apj, 670, 1283

\bibitem[{{Ohsuga} \& {Mineshige}(2011)}]{ohsuga_mineshige2011}
{Ohsuga} K., {Mineshige} S., 2011, \apj, 736, 2

\bibitem[{{Ohsuga} {et~al}\mbox{.}(2009){Ohsuga}, {Mineshige}, {Mori}, \&
  {Kato}}]{ohsuga+2009}
{Ohsuga} K., {Mineshige} S., {Mori} M., {Kato} Y., 2009, \pasj, 61, L7

\bibitem[{{Ohsuga} {et~al}\mbox{.}(2005){Ohsuga}, {Mori}, {Nakamoto}, \&
  {Mineshige}}]{ohsuga+05}
{Ohsuga} K., {Mori} M., {Nakamoto} T., {Mineshige} S., 2005, \apj, 628, 368

\bibitem[{{Omukai}(2001)}]{omukai_2001}
{Omukai} K., 2001, \apj, 546, 635

\bibitem[{{Pacucci} \& {Ferrara}(2015)}]{2015MNRAS.448..104P}
{Pacucci} F., {Ferrara} A., 2015, \mnras, 448, 104

\bibitem[{{Park} \& {Ricotti}(2011)}]{Park_Ricotti_2011}
{Park} K., {Ricotti} M., 2011, \apj, 739, 2

\bibitem[{{Park} \& {Ricotti}(2012)}]{Park_Ricotti_2012}
{Park} K., {Ricotti} M., 2012, \apj, 747, 9

\bibitem[{{Pezzulli}, {Valiante} \& {Schneider}(2016){Pezzulli}, {Valiante}, \&
  {Schneider}}]{Pezzulli_2016}
{Pezzulli} E., {Valiante} R., {Schneider} R., 2016, \mnras, 458, 3047

\bibitem[{{Regan}, {Johansson} \& {Haehnelt}(2014){Regan}, {Johansson}, \&
  {Haehnelt}}]{2014MNRAS.439.1160R}
{Regan} J.~A., {Johansson} P.~H., {Haehnelt} M.~G., 2014, \mnras, 439, 1160

\bibitem[{{Ryu} {et~al}\mbox{.}(2016){Ryu}, {Tanaka}, {Perna}, \&
  {Haiman}}]{Ryu_2016}
{Ryu} T., {Tanaka} T.~L., {Perna} R., {Haiman} Z., 2016, \mnras, 460, 4122

\bibitem[{{Sakurai}, {Inayoshi} \& {Haiman}(2016){Sakurai}, {Inayoshi}, \&
  {Haiman}}]{Sakurai_2016}
{Sakurai} Y., {Inayoshi} K., {Haiman} Z., 2016, \mnras, 461, 4496

\bibitem[{{Sakurai} {et~al}\mbox{.}(2017){Sakurai}, {Yoshida}, {Fujii}, \&
  {Hirano}}]{Sakurai_2017}
{Sakurai} Y., {Yoshida} N., {Fujii} M.~S., {Hirano} S., 2017, ArXiv e-prints

\bibitem[{{S{\c a}dowski} {et~al}\mbox{.}(2015){S{\c a}dowski}, {Narayan},
  {Tchekhovskoy}, {Abarca}, {Zhu}, \& {McKinney}}]{Sadowski_2015}
{S{\c a}dowski} A., {Narayan} R., {Tchekhovskoy} A., {Abarca} D., {Zhu} Y.,
  {McKinney} J.~C., 2015, \mnras, 447, 49

\bibitem[{{Shakura} \& {Sunyaev}(1973)}]{SS_1973}
{Shakura} N.~I., {Sunyaev} R.~A., 1973, \aap, 24, 337

\bibitem[{{Silk} \& {Rees}(1998)}]{Silk_Rees_1998}
{Silk} J., {Rees} M.~J., 1998, \aap, 331, L1

\bibitem[{{Soltan}(1982)}]{Soltan_1982}
{Soltan} A., 1982, \mnras, 200, 115

\bibitem[{{Stone}, {K{\"u}pper} \& {Ostriker}(2017){Stone}, {K{\"u}pper}, \&
  {Ostriker}}]{Stone_2017}
{Stone} N.~C., {K{\"u}pper} A.~H.~W., {Ostriker} J.~P., 2017, \mnras, 467, 4180

\bibitem[{{Sugimura} {et~al}\mbox{.}(2016){Sugimura}, {Hosokawa}, {Yajima}, \&
  {Omukai}}]{sugimura+16}
{Sugimura} K., {Hosokawa} T., {Yajima} H., {Omukai} K., 2016, arXiv: 1610.03482

\bibitem[{{Sugimura}, {Omukai} \& {Inoue}(2014){Sugimura}, {Omukai}, \&
  {Inoue}}]{sugimura_2014}
{Sugimura} K., {Omukai} K., {Inoue} A.~K., 2014, \mnras, 445, 544

\bibitem[{{Takahashi} {et~al}\mbox{.}(2016){Takahashi}, {Ohsuga}, {Kawashima},
  \& {Sekiguchi}}]{takahashi2016}
{Takahashi} H.~R., {Ohsuga} K., {Kawashima} T., {Sekiguchi} Y., 2016, \apj,
  826, 23

\bibitem[{{Tanaka} \& {Haiman}(2009)}]{TH09}
{Tanaka} T., {Haiman} Z., 2009, \apj, 696, 1798

\bibitem[{{Valiante} {et~al}\mbox{.}(2016){Valiante}, {Schneider}, {Volonteri},
  \& {Omukai}}]{Valiante_2016}
{Valiante} R., {Schneider} R., {Volonteri} M., {Omukai} K., 2016, \mnras, 457,
  3356

\bibitem[{{Visbal}, {Haiman} \& {Bryan}(2014){Visbal}, {Haiman}, \&
  {Bryan}}]{2014MNRAS.445.1056V}
{Visbal} E., {Haiman} Z., {Bryan} G.~L., 2014, \mnras, 445, 1056

\bibitem[{{Volonteri}, {Haardt} \& {Madau}(2003){Volonteri}, {Haardt}, \&
  {Madau}}]{VHM03}
{Volonteri} M., {Haardt} F., {Madau} P., 2003, \apj, 582, 559

\bibitem[{{Volonteri} \& {Rees}(2005)}]{VR_2005}
{Volonteri} M., {Rees} M.~J., 2005, \apj, 633, 624

\bibitem[{{Watarai} {et~al}\mbox{.}(2000){Watarai}, {Fukue}, {Takeuchi}, \&
  {Mineshige}}]{watarai+00}
{Watarai} K.-y., {Fukue} J., {Takeuchi} M., {Mineshige} S., 2000, \pasj, 52,
  133

\bibitem[{{Whalen} \& {Norman}(2006)}]{Whalen_2006}
{Whalen} D., {Norman} M.~L., 2006, \apjs, 162, 281

\bibitem[{{Whalen} \& {Norman}(2008)}]{Whalen_2008}
{Whalen} D.~J., {Norman} M.~L., 2008, \apj, 672, 287

\bibitem[{{Wolcott-Green}, {Haiman} \& {Bryan}(2011){Wolcott-Green}, {Haiman},
  \& {Bryan}}]{WHB_2011}
{Wolcott-Green} J., {Haiman} Z., {Bryan} G.~L., 2011, \mnras, 418, 838

\bibitem[{{Wolcott-Green}, {Haiman} \& {Bryan}(2017){Wolcott-Green}, {Haiman},
  \& {Bryan}}]{WHB_2017}
{Wolcott-Green} J., {Haiman} Z., {Bryan} G.~L., 2017, \mnras

\bibitem[{{Wu} {et~al}\mbox{.}(2015){Wu}, {Wang}, {Fan}, {Yi}, {Zuo}, {Bian},
  {Jiang}, {McGreer}, {Wang}, {Yang}, {Yang}, {Thompson}, \&
  {Beletsky}}]{Wu_2015}
{Wu} X.-B. {et~al.}, 2015, \nat, 518, 512

\bibitem[{{Yajima} \& {Khochfar}(2016)}]{2016MNRAS.457.2423Y}
{Yajima} H., {Khochfar} S., 2016, \mnras, 457, 2423

\bibitem[{{Yu} \& {Tremaine}(2002)}]{Yu_Tremaine_2002}
{Yu} Q., {Tremaine} S., 2002, \mnras, 335, 965

\end{thebibliography}
}
\end{document}